\documentclass[conference]{IEEEtran}
\usepackage{amsmath}
\usepackage[usenames,dvipsnames,svgnames,table]{xcolor}
\usepackage{url}
\usepackage{graphicx}
\usepackage[utf8]{inputenc}
\usepackage{booktabs}
\usepackage{colortbl}

\begin{document}

\title{HPA: An Opportunistic Approach to Embedded Energy Efficiency}

\author{\IEEEauthorblockN{Baptiste Delporte and Roberto Rigamonti and Alberto Dassatti}
  \IEEEauthorblockA{Reconfigurable and Embedded Digital Systems Institute\\
    REDS HEIG-VD, School of Business and Engineering Vaud\\
    HES-SO, University of Applied Sciences Western Switzerland\\
    Email: name.surname@heig-vd.ch}}

\maketitle

\begin{abstract}
  Reducing energy consumption is a challenge that is faced on a daily basis by teams from the High-Performance Computing
  as well as the Embedded domain.
  This issue is mostly attacked from an hardware perspective, by devising architectures that put energy efficiency as a primary target,
  often at the cost of processing power.
  Lately, computing platforms have become more and more heterogeneous, but the exploitation of these additional capabilities is so complex
  from the application developer's perspective that they are left unused most of the time, resulting therefore in a supplemental
  waste of energy rather than in faster processing times.

  In this paper we present a transparent, on-the-fly optimization scheme that allows a generic application to automatically exploit the available
  computing units to partition its computational load.
  We have called our approach Heterogeneous Platform Accelerator (HPA).
  The idea is to use profiling to automatically select a computing-intensive candidate for acceleration, and then
  distribute the computations to the different units by off-loading blocks of code to them.

  Using an NVIDIA Jetson TK1 board, we demonstrate that not only HPA results in faster processing speed, but also in a considerable reduction
  in the total energy absorbed.
\end{abstract}

\section{Introduction}
The energy consumption problem is common to the whole Computer Science world, as it is one of the major limiting factor to more powerful devices
--- especially embedded ones, which would otherwise run out of battery very quickly~\cite{Schmitz04} ---
as well as one of the major sources of expenses (and pollution) for big data centers~\cite{Koomey08,Ruth09}.
Solutions to it range from accepting a performance reduction in return for longer battery life, as in the Intel Atom processor~\cite{Beavers09}, to radical relocations of large
data centers in cold regions~\cite{Hancock09}.

\begin{figure}
  \centering
    \includegraphics[width=.9\columnwidth]{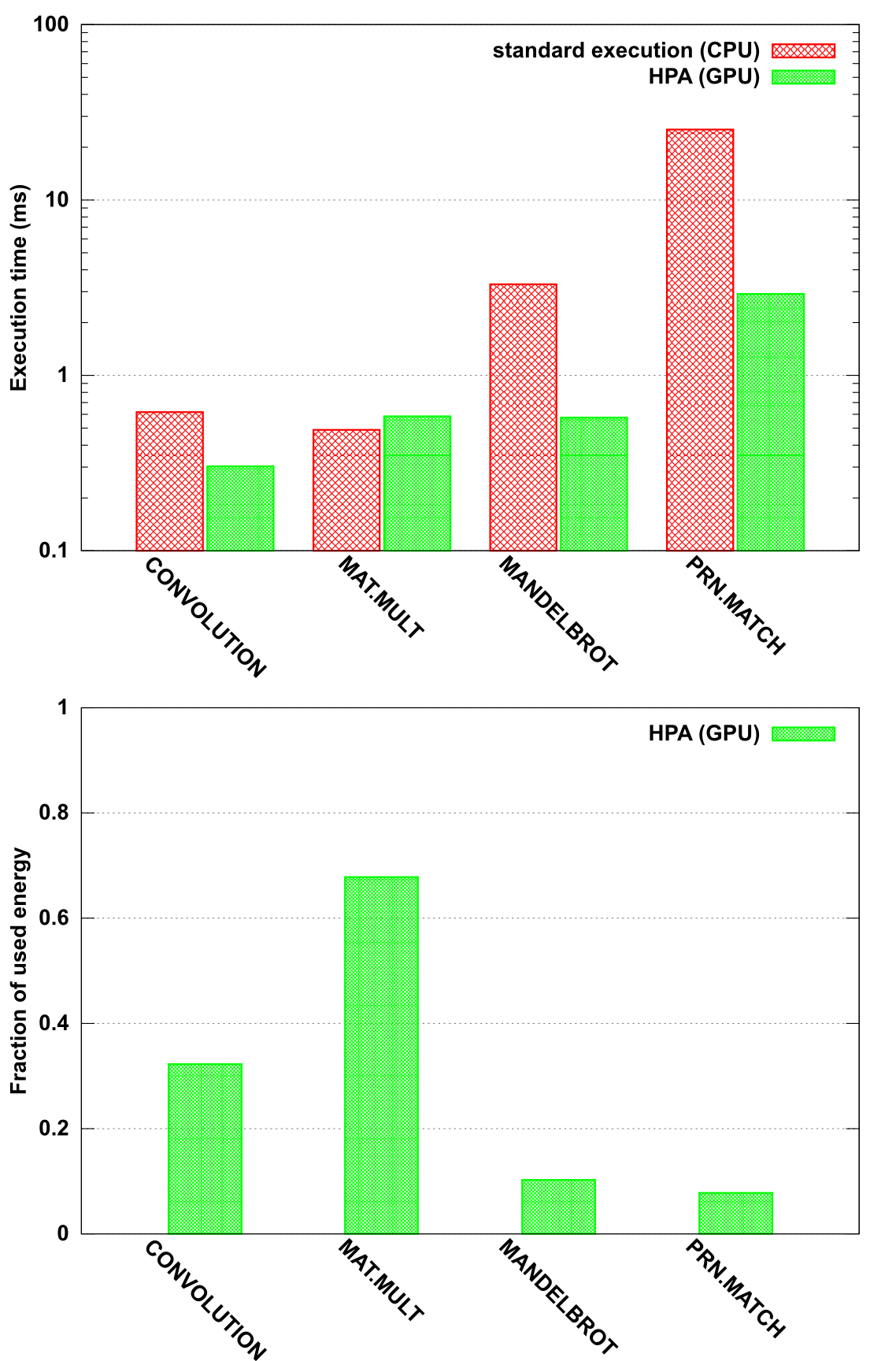}
  \caption{{\bf (top)} Execution time (in milliseconds over a logarithmic scale) of four test
    algorithms in the two situations we consider: standard CPU execution, and execution on the GPU
    in the context of our framework. We can see that HPA largely outperforms CPU execution in all
    but one test, MAT.MULT.
    However, if we consider the fraction of energy absorbed with respect to standard CPU execution {\bf (bottom)},
    we see that the execution in the context of HPA is considerably more efficient even when it takes more time to
    complete the computations.}
  \label{fig:algos}
\end{figure}

In parallel to this drift towards greener forms of computing, the last few years saw a general trend in the direction of heterogeneous computing platforms, mostly issued from the
acknowledgement of the limitations imposed by physics and technology to the pursuit of ever faster devices~\cite{Khan11,Chung10,Kumar05}.
While having multiple, more energy-efficient accelerators on the same board --- or even the same chip --- should have led to increased performances and reduced power absorption
(as each unit would be used in an optimal way, for instance, highly-parallel units for highly-parallel tasks),
reality shows that the former target is only attained for very specific software customized for the accelerator, while the latter is rarely achieved due to the presence of a largely unused ---
but often leaking power --- resources.
The problem this time lies on the software side: writing software for an inhomogeneous, ever changing set of targets is difficult, expensive, and requires skilled and motivated developers and
constant maintenance.
This fact contrasts with the very same nature of software, which usually evolves at a slow pace~\cite{Brooks95}, and has heavily influenced current programming styles;
As an example, Android developers are accustomed to not knowing on which platform the software they are writing is going to be executed.
Indeed, most of the optimizations in JAVA, C\#, Swift and other widespread programming languages are left at run-time, when the code is first execute and information about the target are
available, generating in this way an optimized bytecode.
However, this strategy reduces the scope of the optimizations available to the developer, forcing him to produce a generic code that is unlikely to fully exploit hardware capabilities.
To make software capable of dealing with an heterogeneous environment and broader improvement opportunities, a reasonable approach seems to be automation~\cite{Damschen15a,Damschen15b,Delporte15a}.

In this paper we present an optimization system, called Heterogeneous Platform Accelerator (HPA), that automatically detects and delegates computing-intensive tasks to a dedicated accelerator,
taking as an example of such accelerator the GPU chip available on an NVIDIA Jetson TK1 board.
Our system grounds on the LLVM's Just-In-Time (JIT) compiler MCJIT~\cite{Lattner04,Lattner08,Lattner11}: executed code chunks are periodically analyzed by the perf\_event~\cite{Weaver13}
performance monitor and, once a particular function is classified as computing-intensive, it is ran through the Polly tool~\cite{Grosser12} to investigate whether it is parallelizable.
If this is the case, the function is off-loaded to the GPU.
The big advantage of this approach is that everything is totally transparent to the developer: he does not have to know on which platform his software will be deployed, as the
potential accelerators will be discovered at run-time, nor he has to add any markings or adopt a particular library to allow the optimization.
Moreover, the optimization is dynamic: should the performance analyzer detect that the GPU is slower than the CPU for a given task, or that it is desirable to leave the GPU
to another function that would make better use of it, the system can revert its operations and recall back the currently off-loaded code.
The unique limitation to the strategy we propose is the amount of intelligence and previous knowledge we are willing to put in the optimization scheme.

\section{Related Work}
In the last few years a remarkable increase in the interest about heterogeneous platforms has been observed; GPUs, by their nature and capabilities, have been deep in the center of this
small revolution.
Whilst a large number of proposals have focused, with good results, on devising a language layer that could ease the development on such platforms~\cite{Danalis10,Stone10,Kyriazis13},
relatively few implementations that automatically perform a multi-target optimization appeared.

An interesting approach, called BAAR, is presented in~\cite{Damschen15a,Damschen15b}.
The code to be executed is at first statically analyzed using Polly~\cite{Grosser12}, a state-of-the-art polyhedral optimizer for automatic parallelization.
If Polly detects that a function deserves being optimized, its code is compiled with the Intel's compiler and off-loaded to an Intel Xeon Phi board.
All data transfers are dealt with using a software layer that handles them using MPI.
While captivating results are given, the major drawback of this approach is that it lacks workload adaptation: the analysis is performed at application's startup, and it does not account for
changes in the context of execution.
The functions to be off-loaded are selected according to a metric that accounts for the number and type of operations to be performed, but this is only a rough indicator of the relative
weight of the function in the context of the whole program's execution.
As a consequence, this strategy is not reactive with respect to the user input, and this could lead to suboptimal choices ---
consider, for instance, the multiplication of a matrix whose size is user-specified.

Another proposal, named StarPU~\cite{Augonnet11}, provides an API and a pragma-based environment that, coupled with a run-time scheduler
for heterogeneous hardware, composes a complete solution.
While the main focus of the project is CPU/GPU systems, it could be extended to less standard systems.
However, the developer is requested to learn a new API and foresee which parts of the code should be optimized.
This analysis is dependent on both the input and the available resources and thus, despite the additional efforts required to the programmers, will not always lead to optimal results.

An approach which is not restricted to a single target type is SoSOC~\cite{Nasrallah13}: it consists of a library that presents a friendly interface to the programmer
and allows functions to be dispatched to a set of targets based on either the developer's wishes or some statistics computed during early runs.
This solution adds some dynamicity with respect to StarPU, but the developer not only has to learn (yet) another library, but also someone has to provide handcrafted code
for any specialized unit of interest.
This is a considerable waste of time and resources, and limits the applicability of the system to the restricted subset of architectures directly supported by the development team.

VPE~\cite{Delporte15a} represents an evolution of SoSOC: its focus is on transparency, which means that the developer does not even have to be aware that his code is going to be accelerated.
Starting by executing the code in an LLVM's JIT-based framework, VPE detects which functions are computing-intensive, and off-loads them to a remote unit.
The code deployed is the very same code that is executed on the CPU, therefore no effort is requested to develop custom implementations.
Results on a board based on the TI-DM3730 chip, which features a C64+ DSP processor, show gains up to $32\times$ in performance.
Nonetheless, the implementation is heavily customized for the examined board, since LLVM does not provide a backend for this platform, and therefore a set of scripts has to compile
ahead-of-time the functions using the TI proprietary compiler.
When a candidate function is found, VPE off-loads its operations to the DSP by executing the previously-compiled code on the data that have been allocated in a shared memory region.
In this paper we adopt a similar approach, but with some relevant differences.
Besides the different target family --- a GPU --- and the availability of a backend --- that allows us to compile on-the-fly the portions of code of interest ---, we put a
strong bias towards energy efficiency, investigating how the optimization impacts on the power absorption.
Moreover, we introduce an additional step in which computing-intensive functions are analyzed to detect the possibility of parallelizing them: indeed, it would be of little value to
off-load sequential code to the 192-cores GPU of the Jetson TK1 board.
Finally, we quantitatively investigate how the optimizations allowed by the use of a JIT-compiler in place of ahead-of-time compilation impact, in the context of our optimization scheme,
both processing speed and power consumption.

\section{Proposed Approach}
The fundamental building block of our proposal is the LLVM's JIT compiler.
LLVM is an alternative to the widely known GCC compiler with a neater separation between the front-end, the optimization, and the back-end steps~\cite{Lattner08}.
This separation is largely due to the adoption of an enriched assembly language, called Intermediate Representation (IR), acting as a shared language between the different stages~\cite{Lattner04}.
LLVM's community is very active, and recently a new JIT engine, called ORC, has been proposed.
The previous engine, called MCJIT, presents indeed a serious issue: it has been designed to operate on modules --- that is, aggregates of functions ---, and once a module has been finalized
(a mandatory step for execution), it is not possible to modify its code anymore.
ORC solves this problem by allowing on-the-fly changes; however, it is currently under development and available for the x86\_64 architecture only.
For this reason, we have adopted the old JIT interface and implemented the transition across accelerators using a caller mechanism akin to that of~\cite{Delporte15a}.
In particular, we analyze at application startup which functions are available in the code --- we automatically detect and exclude all system calls and I/O-based functions
from our analysis, as we cannot optimize them with out approach  ---, and we replace function invocations with a caller that, when the function
is not off-loaded, simply executes the desired function on the CPU via a function pointer.
Once a function is selected for off-loading, we alter the function pointer to make it point to the function ready to be executed on the GPU.
Please note that this caller overhead will be removed once ORC will be released for a broader selection of platforms.

\begin{figure*}[t]
  \centering
\includegraphics[width=\linewidth]{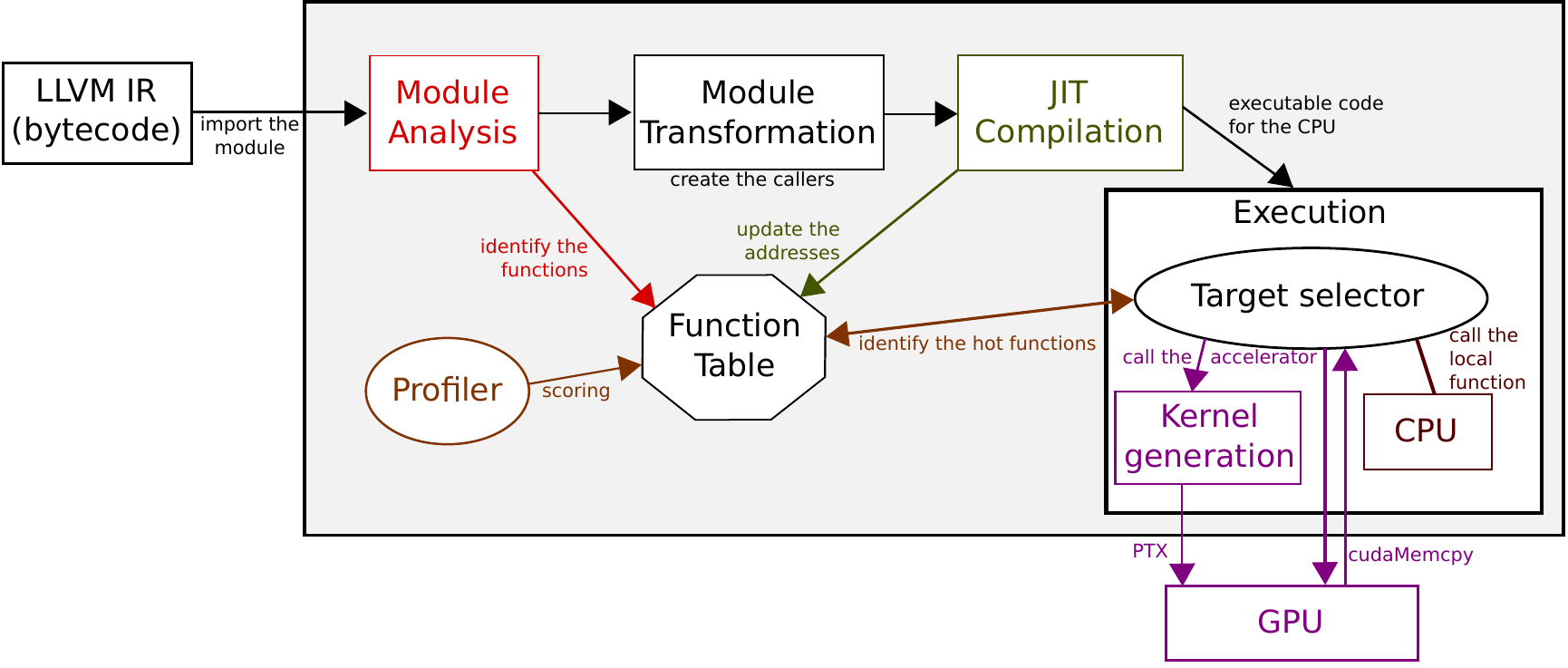}
\caption{Architecture of the HPA framework. An input program, in LLVM's IR bytecode format, is at first analyzed.
  Then, functions are identified and callers are created for all the potential candidates to off-loading.
  After the JIT-compilation step, the code is executed and profiling information are acquired.
  Once HPA detects that a given function is worth off-loading to the GPU, it transfers the required data and loads (or generates, if this is the first time the function is invoked) the PTX code
  that will be executed on the accelerator.}
\label{fig:arch}
\end{figure*}

Detecting whether a function deserves off-loading on the GPU or not is a two-step process:
at first, functions that perform heavy computations are identified by using perf\_event~\cite{Weaver13}.
perf\_event collects very detailed statistics about software and hardware counter, and permits us to easily identify performance bottlenecks.
In the context of this paper we rely on the CPU usage alone to identify functions which might benefit from being off-loaded, but many more optimizations could be devised;
For instance, inspecting memory access patterns could suggest an improved layout for a data structure that reduces cache misses~\cite{Chilimbi99}.
After having obtained a sorted list of functions candidate for acceleration, we inspect each of them sequentially, checking with Polly~\cite{Grosser12} whether it is parallelizable or not.
Polly operates at the IR level and starts by translating the code to optimize to a polyhedral representation~\cite{Bondhugula08}.
It then detects the Static Controls Parts (SCoPs)~\cite{Damschen15a} of the code matching a specific canonical form, and via a set of LLVM passes it performs the
optimization and detects parallelism.
In particular, it places OpenMP~\cite{Dagum98} markings around parallelizable blocks of code.
We detect those markings and use them to guide our choices.
For the sake of simplicity, we operate at function level, off-loading entire functions to the accelerator.
Our approach is, however, independent with respect to the choice of the scale, so we could as well operate at the basic-block level~\cite{Lattner04} --- which could be interesting,
for instance, for a multi-threaded function.

Once we have selected a function to off-load to the GPU, we generate on-the-fly the PTX code~\cite{Nvidia08} that is sent to the GPU by using the LLVM's backend.
Data are transferred to an accessible memory region using the dedicated CUDA instructions, and the execution is started.
The results are finally transferred back once the computation is finished.
Although CUDA 6.5 offers an Unified Memory scheme~\cite{Harris13}, we experimentally verified that its performances are considerably worse than individual cudaMemcpy()
operations.
We have therefore opted for a manual transfer of the parameters, taking the transfer time into account in our measurements.
Please note that, once the memory sharing will be a viable option, the performance of HPA will be further improved.

The architecture of HPA is depicted by Fig.~\ref{fig:arch}.

\section{Results}

We validate our proposal using an NVIDIA Jetson TK1 board, which features a 4+1 ARM Cortex-A15 32-bit processor with a 192-cores GK20A (Kepler) 852MHz GPU.
This GPU, with respect to its predecessors, has a particular focus on energy efficiency.
The installed Linux distribution is a Ubuntu 14.04 with a patched kernel distributed in the L4T (Linux for Tegra) package.
We have set the CPU power governor to ``performance'' to guarantee the CPU the best possible performance.
All the code we use has been compiled using Clang with strong optimization turned on (\emph{-O3}).

We perform experiments with a test set similar to the one used in~\cite{Delporte15a}:
we considered a set of four algorithms inspired by the Computer Language Benchmarks Game\footnote{\url{http://benchmarksgame.alioth.debian.org}}, namely
2D convolution with a square kernel matrix (CONVOLUTION), multiplication of two square matrices (MAT.MULT), Mandelbrot set generation (MANDELBROT),
and search of a nucleotidic pattern in an input DNA sequence (PRN.MATCH).
Contrary to~\cite{Delporte15a}, given the nature of the adopted accelerator, we put no limitations on the use of floating point numbers.
To provide a fair comparison, we fixed the amount of data to process and manually parallelized the CPU implementation of the algorithms to fully exploit the available cores.

To estimate power consumption,
we attached a shunt resistor to the ground power line and performed current measurements by using a Tektronix MSO5104B Mixed Signal Oscilloscope.
We recorded the voltage over extensive periods and then computed the average and standard deviation values for each explored case.

\begin{figure}
  \centering
\includegraphics[width=.9\columnwidth]{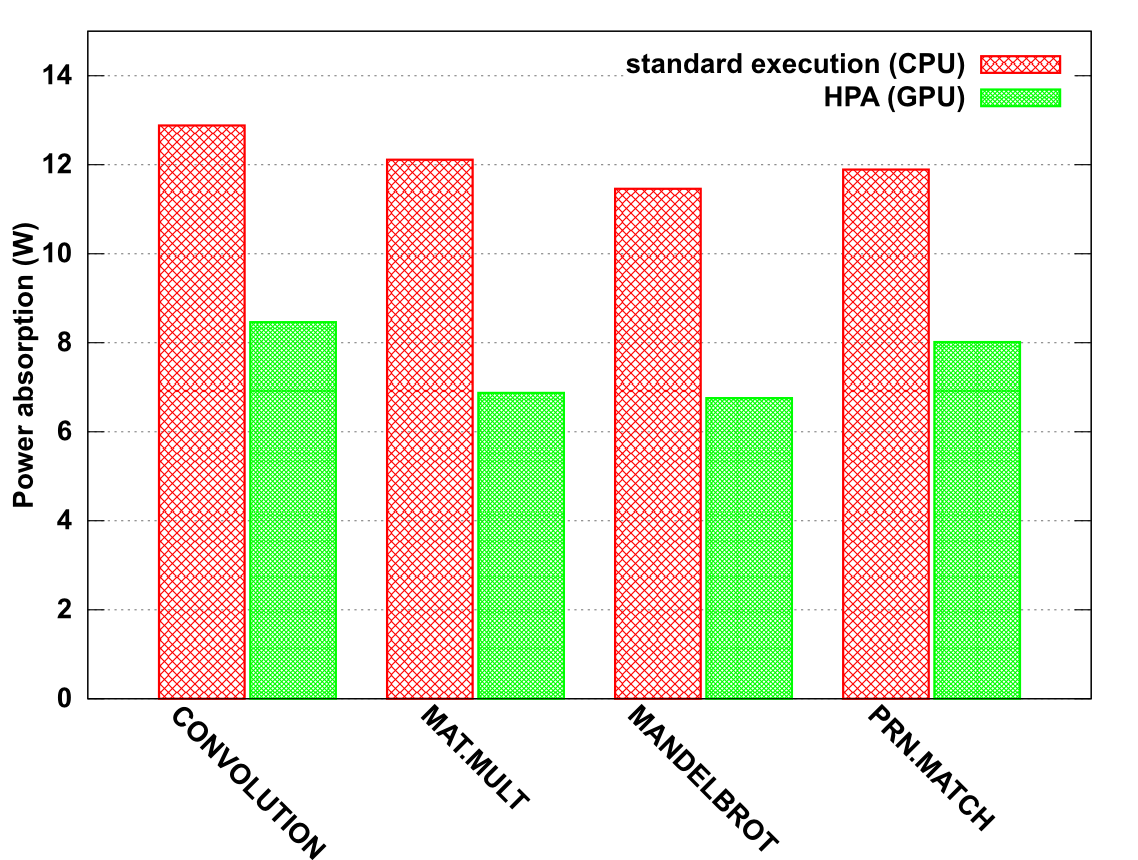}[t]
  \caption{Power absorption in the case of the four considered algorithms for both standard CPU execution and execution on the GPU
    in the context of our framework.}
\label{fig:power}
\end{figure}

Figure~\ref{fig:algos}(top) shows the time required for the execution of each test in the two situations we consider, that is, standard CPU execution and execution on the GPU in the HPA framework.
Executing on the GPU, even if the code is automatically generated and not hand-crafted for it, results in higher performance in all but the MAT.MULT case.
This latter case could be due to a non-GPU-friendly implementation of the standard algorithm, and exposes the major weakness of our approach:
as the code we execute on the accelerator was initially conceived for a standard CPU, we have little hope of achieving the performance of a carefully engineered CUDA algorithm.
In our opinion, however, this drawback is largely offset by the complete transparency we offer the developer.
If the system detects a lower-than-expected performance, it can at any time revert its choices and obtain as worst-case scenario the same performance the original code was supposed to achieve.
Moreover, if we consider the power absorption depicted by Fig.~\ref{fig:power}, we see that, in our experiments, HPA is more energy efficient in all the cases, even
when it takes longer to get to the final result as it is the case for MAT.MULT.
Therefore, we could easily integrate an external policy that dictates which option to favor the most --- best performance, best energy efficiency, or a combination of the two ---
and adapt the algorithm's behavior according to the run-time measurements.
The combined effects of shorted processing times and lower power consumption are depicted by Fig.~\ref{fig:algos}(bottom), where the energy consumption of HPA is normalized over the consumption
in the standard CPU case, while detailed results about execution time and power absorption are reported in Tab.~\ref{tab:algos}.

\begin{figure}
  \centering
\includegraphics[width=.9\columnwidth]{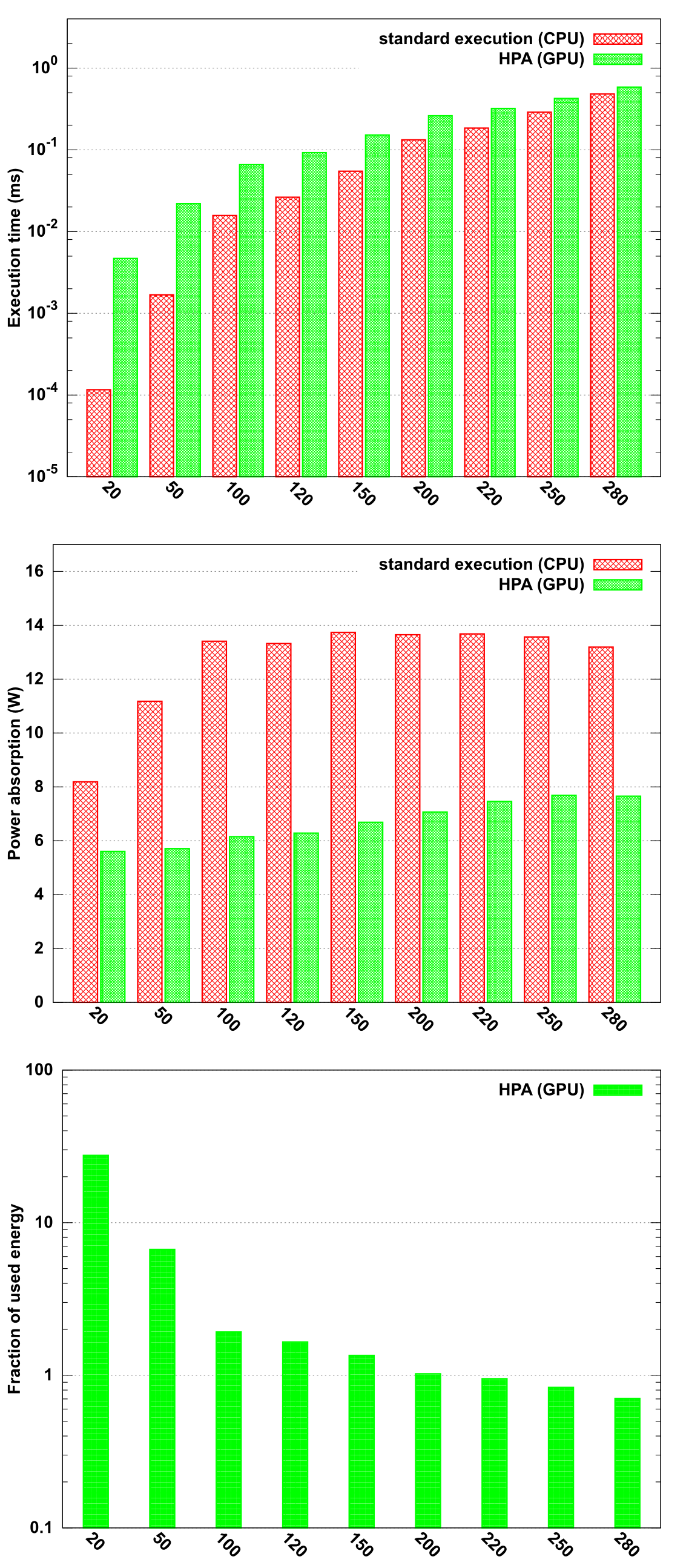}
\caption{{\bf (top)} Execution time, {\bf (center)} power absorption, and {\bf (bottom)} fraction of energy with respect to the standard CPU execution for the MAT.MULT test
  as a function of matrix size.
  The graphs shows that it is orders of magnitude slower to execute the multiplication on the GPU when small matrices are involved;
  Indeed, both the data transfer time and the profiler's overhead exceed any potential gain that could derive by performing the multiplication on the GPU.
  Operating on the GPU becomes interesting, at least from an energetic stance, for matrices bigger than $200 \times 200$.}
\label{fig:m2_mat}
\end{figure}

We have further investigated the MAT.MULT experiment to see the behavior of the system as a function of matrix size.
Figure~\ref{fig:m2_mat} depicts the execution time, the power consumption and the fraction of energy power used by the GPU with respect to standard CPU execution for a large set of matrix sizes.
We can see that it is more convenient to operate on the CPU for matrices smaller than $200 \times 200$.
Indeed, the data transfer we perform and the overhead introduced by the performance analyzer
exceed by several orders of magnitude the computation time, and therefore it can be up to $30\times$ more expensive, in energetic terms, to perform these multiplications on the GPU.
For bigger sizes, despite the slightly higher execution time, the reduced power absorption makes the HPA alternative interesting.

\begin{figure}
  \centering
  \includegraphics[width=.9\columnwidth]{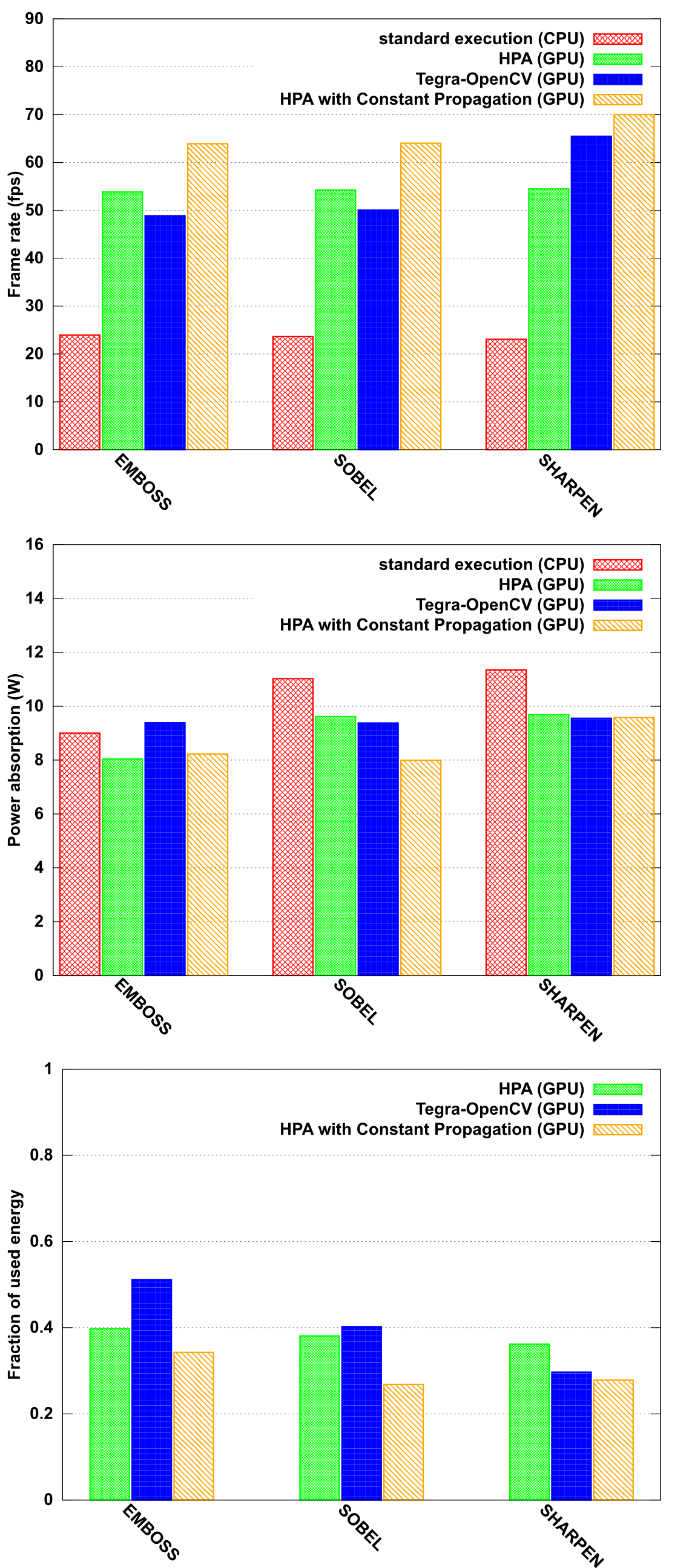}
  \caption{{\bf (top)} Execution time, {\bf (center)} power absorption, and {\bf (bottom)} fraction of energy with respect to the standard CPU execution for the
    image processing demo.
    Three different filter types are considered, each with a different sparsity degree.
    We can see that the Tegra-OpenCV implementation can partially exploit the sparsity, but not as much as the JIT-optimized version --- which is unsurprising, as the
    OpenCV version still has to perform all the multiplications, while the JIT version drops all unnecessary operations.}
\label{fig:muu}
\end{figure}

\begin{figure}
  \centering
  \includegraphics[width=\columnwidth]{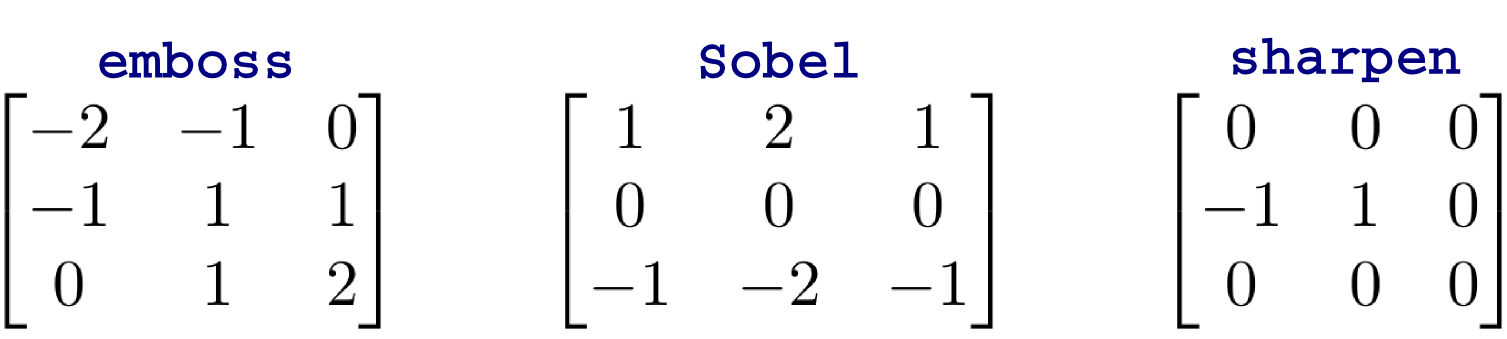}
\caption{Structure of the three filters used in the image processing demo.
  We have chosen them as they present a different degree of sparsity, and we wanted to investigate the capabilities of a JIT-based framework to exploit this information
  when optimizing the computations.}
\label{fig:mat_struct}
\end{figure}

\begin{figure*}
  \centering
  \includegraphics[width=\linewidth]{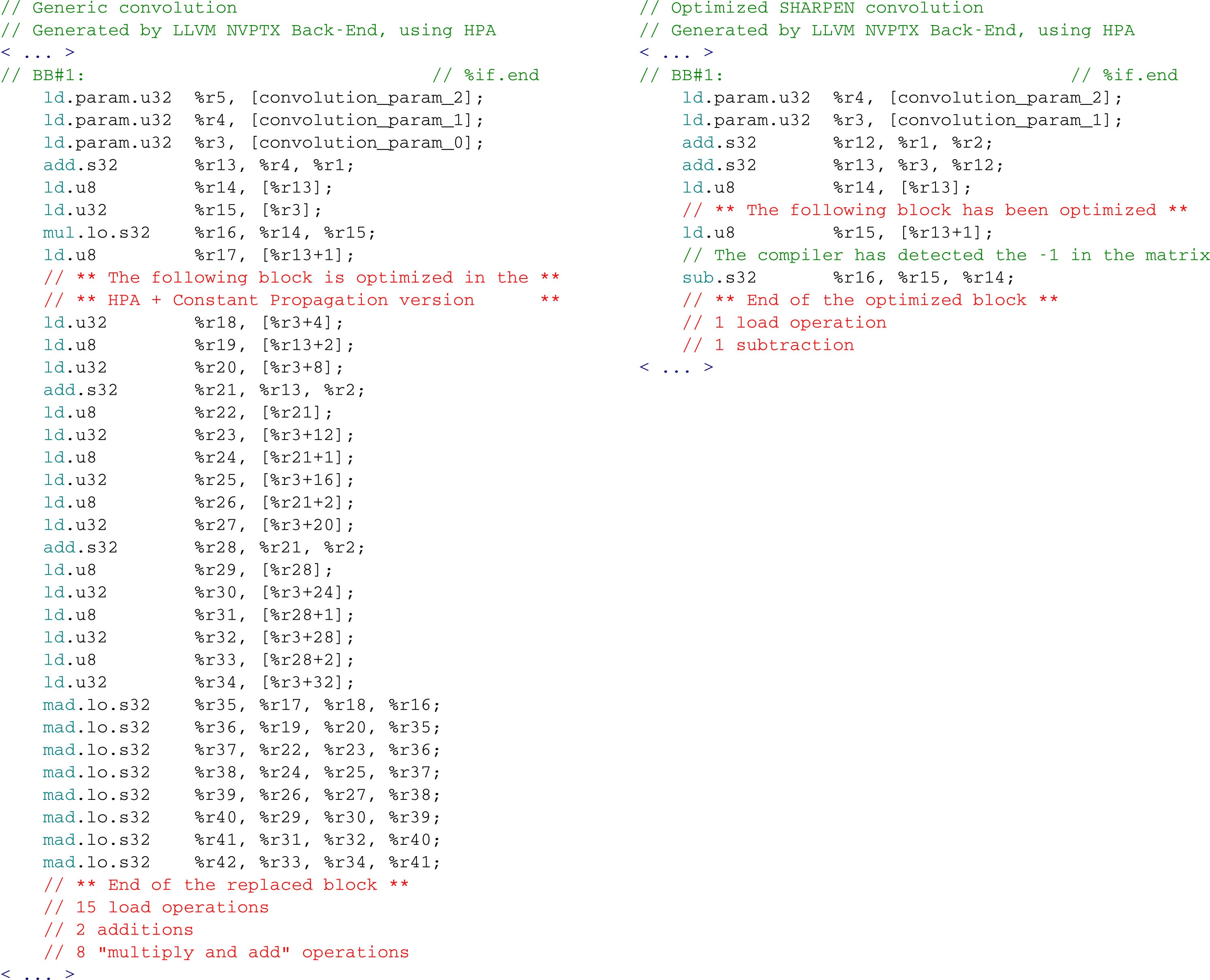}
  \caption{PTX code generated by the LLVM back-end for the generic convolution code {\bf (left)} and the version optimized by
    the JIT Constant Propagation optimization step {\bf (right)}.
    The number of executed instruction is greatly reduced due to the particular shape of the adopted image kernel.}
  \label{fig:code}
\end{figure*}

We then developed an image-processing demo where different 2D filters are applied on the frames of a $1280\times 720$ MPEG-4-encoded video.
The image decompression and visualization tasks are performed using the Tegra version of the OpenCV library to avoid saturating the CPU with operations
unrelated to the image-processing task itself, though this slightly penalizes HPA as it reduces the GPU resources available for the convolutions.
In the comparison, we also considered the convolution operation performed by using the Tegra-OpenCV library: this version of the convolution is indeed optimized
for the accelerator, and we expected it to outperform the HPA version.
Moreover, we wanted to check whether using a JIT compiler --- and thus being capable of adapting the code on-the-fly to the input data --- was giving us some added value.
We therefore JIT-compiled the module including the Constant Propagation optimization, and performed our measures using the three filters depicted by Fig.~\ref{fig:mat_struct}
(\emph{emboss}, \emph{Sobel y-}derivative, and \emph{sharpen}) that exhibit different structures --- in particular, different sparsity.
The results are reported in Fig.~\ref{fig:muu} and detailed in Tab.~\ref{tab:ip}.
Unsurprisingly, operating on the GPU resulted in significantly faster computations and thus, given a lower power absorption, a considerable reduction in the required energy
with respect to CPU execution.
The comparison with the Tegra-OpenCV implementation of the convolution operator is interesting: we can see that the frame rate of the hand-optimized convolution operator
is lower than the one achieved by HPA for both the emboss and the Sobel filter.
This could be due to the JIT compiler performing some small optimizations, and the genericity of the OpenCV version of the convolution preventing it from adopting
architecture-dependent shortcuts.
Also, we can see that, in the sharpening filter case, the OpenCV version can partially benefit of the presence of several zero entries in the filter matrix, but still it cannot outperform
the HPA results when Constant Propagation is used in the JIT compilation.
This latter, can indeed take advantage of the presence of elements with opposed sign, such as $1$ and $-1$, and exploit this information to further speed-up computations.
An example of these optimizations can be seen in Fig.~\ref{fig:code}, which has on the left side the generic convolution code that is normally executed by HPA and on the right side
the code produced by JIT's Constant Propagation in the sharpening filter's case.
These results support our claim that higher efficiency, in both computational and energetic terms, can be achieved by casting a problem in the HPA-JIT framework.

\begin{table*}[t]
  \centering
  \caption{Execution time (in ms) and power absorption (in W) for the four considered algorithms.
    The number reported after the $\pm$ sign represents one standard deviation.
    With ``standard execution (CPU)'' we indicate the plain execution of the algorithm on the four cores of the ARM CPU with no performance collection undergoing,
    while with ``HPA (GPU)'' we indicate the very same code but running on the GPU in the HPA framework}
  \label{tab:algos}
  \begin{tabular}{@{} lccccc @{}}
    \toprule
    {\bf Algorithm}                       & \multicolumn{2}{c}{{\bf standard execution (CPU)}} & \hspace{0.1em} & \multicolumn{2}{c}{{\bf HPA (GPU)}}   \\
                                            \cline{2-3}                                                         \cline{5-6}
                                          & Exec. time [ms]   & Power cons. [W]                &                & Exec. time [ms]   & Power cons. [W]   \\
    \midrule
    \textcolor{NavyBlue}{CONVOLUTION}     & $0.61 \pm 0.001$  & $12.88 \pm 0.008 $             &                & $0.30 \pm 0.043$  & $8.46 \pm 0.029 $ \\
    \textcolor{NavyBlue}{MAT.MULT}        & $0.49 \pm 0.015$  & $12.11 \pm 0.001 $             &                & $0.58 \pm 0.088$  & $6.87 \pm 0.001 $ \\
    \textcolor{NavyBlue}{MANDELBROT}      & $3.30 \pm 0.018$  & $11.46 \pm 0.013 $             &                & $0.57 \pm 0.131$  & $6.75 \pm 0.016 $ \\
    \textcolor{NavyBlue}{PRN.MATCH}       & $25.20\pm 0.008$  & $11.89 \pm 0.005 $             &                & $2.91 \pm 0.280$  & $8.01 \pm 0.006 $ \\
    \hline
  \end{tabular}
\end{table*}

\begin{table*}[t]
  \centering
  \caption{Frame rate (in fps) and power absorption (in W) for the image processing test.
    The number reported after the $\pm$ sign represents one standard deviation.
    With ``standard execution (CPU)'' we indicate the plain execution of the algorithm on the four cores of the ARM CPU with no performance collection undergoing,
    with ``HPA (GPU)'' we indicate the very same code but running on the GPU in the HPA framework, with ``Tegra-OpenCV (GPU)'' we indicate the execution on the GPU of the
    Tegra-OpenCV optimized convolution code, and with ``HPA + Constant Propagation (GPU)'' the HPA version but with the Constant Propagation optimization of the JIT framework active
  }
  \label{tab:ip}
  \begin{tabular}{@{} lcccccc @{}}
    \toprule
    {\bf Execution target}                                 & \multicolumn{3}{c}{{\bf Frame rate [fps]}}             & \multicolumn{3}{c}{{\bf Power absorption [W]}}            \\
                                                           \cline{2-4}                                              \cline{5-7}
                                                           & Emboss           & Sobel            & Sharpen          & Emboss            & Sobel             & Sharpen           \\
    \midrule
    \textcolor{NavyBlue}{standard execution (CPU)}         & $23.97 \pm 0.61$ & $23.68 \pm 0.61$ & $23.07 \pm 0.78$ & $9.00 \pm 0.015$  & $11.02 \pm 0.027$ & $11.35 \pm 0.094$ \\
    \textcolor{NavyBlue}{HPA (GPU)}                        & $53.84 \pm 6.80$ & $54.25 \pm 7.69$ & $54.47 \pm 7.81$ & $8.04 \pm 0.022$  & $9.62 \pm 0.018$  & $9.69 \pm 0.043$  \\
    \textcolor{NavyBlue}{Tegra-OpenCV (GPU)}               & $48.88 \pm 3.12$ & $50.06 \pm 2.91$ & $65.48 \pm 7.15$ & $9.39 \pm 0.213$  & $9.38 \pm 0.093$  & $9.55 \pm 0.009$  \\
    \textcolor{NavyBlue}{HPA + Constant Propagation (GPU)} & $63.92 \pm 8.45$ & $64.02 \pm 9.03$ & $70.02 \pm 7.90$ & $8.22 \pm 0.007$  & $7.99 \pm 0.023$  & $9.58 \pm 0.104$  \\
    \hline
  \end{tabular}
\end{table*}

\section{Conclusion}
In this paper we presented an opportunistic strategy to increase energy efficiency by automatically off-loading computational intensive fragments of easily-parallelizable code
to a GPU accelerator.
As a side effect, we get a significant increment in the overall performances, as not only these computations are performed faster, but also the main CPU load is relieved and
thus it can accomplish further tasks while waiting for the computations to terminate.
We supported our claims with thorough experiments on several algorithms and an image processing task.

As future work we will focus on defining other optimization strategies and investigate platforms with a higher degree of heterogeneity, choosing at run-time the target that is expected
to give the highest energy efficiency or fit best to a set of user-defined policies.

\bibliographystyle{ieeetr}
\bibliography{paper}

\end{document}